\documentclass[12pt]{iopart}
\usepackage{graphics,iopams,mathrsfs}

\makeatletter
\def\@fnsymbol#1{^{\thefootnote}\relax}
\makeatother

\begin{document}

\title[Band filling dependence of the Curie temperature in CrO$_2$]
{Band filling dependence of the Curie temperature in CrO$_2$}

\author{I V Solovyev$^{1,2}$, I V Kashin$^2$, V. V. Mazurenko$^2$}

\address{$^1$ Computational Materials Science Unit,
National Institute for Materials Science,
1-1 Namiki, Tsukuba, Ibaraki 305-0044, Japan}
\ead{solovyev.igor@nims.go.jp}
\address{$^2$ Department of Theoretical Physics and Applied Mathematics,
Ural Federal University,
Mira str. 19, 620002 Ekaterinburg, Russia}

\begin{abstract}
Rutile CrO$_2$ is an important half-metallic ferromagnetic material, which is also widely used in magnetic recording.
In an attempt to find the conditions, which lead to the increase the Curie temperature ($T_{\rm C}$),
we study theoretically the band-filling dependence of interatomic exchange interactions in the rutile compounds.
For these purposes, we use the effective low-energy model for the magnetic $t_{2g}$ bands,
derived from the first-principles electronic structure calculations in the Wannier basis, which is solved
by means of dynamical mean-field theory. After the solution, we calculate the interatomic exchange interactions,
by using the theory of infinitesimal spin rotations, and evaluate $T_{\rm C}$. We argue that,
as far as the Curie temperature is concerned, the band filling realized in CrO$_2$ is far from being the optimal one
and much higher $T_{\rm C}$ can be obtained by decreasing the number of $t_{2g}$ electrons ($n$) via the hole doping.
We find that the optimal $n$ is close to $1$, which should correspond to the case of VO$_2$, provided that it
is crystallized in the rutile structure. This finding was confirmed by using the experimental rutile structure for
both CrO$_2$ and VO$_2$ and reflects the general tendency towards ferromagnetism for the narrow-band compounds at the
beginning of the band filling. In particular, our results suggest that the strong ferromagnetism can be achieved in the
thin films of VO$_2$, whose crystal structure is controlled by the substrate.
\end{abstract}


\maketitle

\section{Introduction}
CrO$_2$ is a rare example of stoichiometric oxide, which demonstrates metallic and ferromagnetic (FM) properties.
It is widely used in magnetic recording and
considered as one of the best particulate ever invented for these purposes~\cite{SkomskiCoey,Skomski}.
Besides magnetorecording, chromium dioxide has attracted a considerable scientific interest due to its
half-metallic electronic structure~\cite{Schwarz,Mazin,Chioncel,HMRevModPhys},
which is intensively studied today due to its implication in various spin-dependent transport phenomena \cite{Singh}.

  In our previous work~\cite{PRB2015}, we have studied the microscopic origin
of the half-metallic ferromagnetism in CrO$_2$ by combining
first-principles electronic structure calculations with the
model Hamiltonian approach and modern many-body methods for
treating the electron-electron correlations. We have demonstrated that
the problem is indeed highly nontrivial: at the first glance,
the ferromagnetism in CrO$_2$ can be easily explained by the Hund's
rule related exchange processes in the narrow $t_{2g}$ band, located near the Fermi level.
However, the electron-electron correlations, rigorously treated in the
frameworks of dynamical mean-field theory (DMFT)~\cite{DMFTRevModPhys}, tend to destabilize
the FM state.
The ferromagnetism in the stoichiometric CrO$_2$ reemerges only if, besides
conventional kinetic energy changes in the $t_{2g}$ band, to
consider other mechanism, involving direct exchange and
magnetic polarization of the oxygen band. We have shown how all these
contributions can be evaluated using first-principles
electronic structure calculations in the local-density approximation (LDA).

\begin{figure}[h!]
\begin{center}
\resizebox{15cm}{!}{\includegraphics{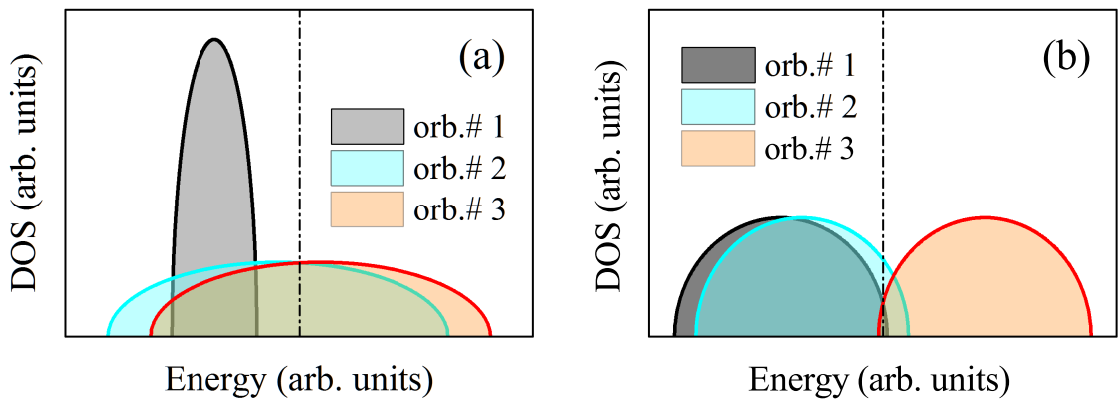}}
\end{center}
\caption{\label{fig.modelDOS} Schematic view on the majority-spin electronic structure of
the transition-metal dioxides with the rutile structure at $n=2$ (where $n$ is the number of electrons
in the $t_{2g}$ band per one transition-metal site), which supports the ferromagnetic ground state (a)
and antiferromagnetic ground state (b). The Fermi level is shown by dot-dashed line.}
\end{figure}

  The main purpose of the present work is to explore the question: How can one further stabilize the
metallic FM state and increase the Curie temperature ($T_{\rm C}$) in CrO$_2$?
This is indeed a very important technological and fundamental problem.
From the technological point of view, $T_{\rm C}$ in CrO$_2$ is not exceptionally high (about $390$ K~\cite{Skomski}),
so that the magnetic properties at the room temperature are deteriorated by the thermal fluctuations.
From the theoretical point of view, the problem is also highly nontrivial because
the robustness of the ferromagnetism in CrO$_2$ strongly depends on
details of the electronic structure.
Indeed, the magnetic properties of CrO$_2$ are predetermined by the behavior
of $t_{2g}$ bands, located near the Fermi level. In the rutile structure, the $t_{2g}$ orbitals belong to
different representations of the point group. Therefore, depending on the crystal distortion,
one can expect rather dissimilar behavior
of these orbital contributions, and hypothetically one can imagine rather
different types of the electronic structure, which will stabilize different magnetic states.
For example, if one of the $t_{2g}$ electrons resides
on the localized orbital, which forms a narrow band below the Fermi level, and
polarizes the itinerant electrons in two other
nearly degenerate bands (see Fig.~\ref{fig.modelDOS}a),
one can expect a strong tendency towards ferromagnetism driven by the
direct exchange (DE) mechanism~\cite{PRL99}.
This scenario for CrO$_2$, which is reminiscent of the physics of colossal magnetoresistive
manganites~\cite{Dagotto}, was proposed in ref.~\cite{Korotin}
and since then was actively discussed in the literature.
From the viewpoint of the DE physics, the electronic structure depicted in Fig.~\ref{fig.modelDOS}a is nearly optimal
for stabilizing the FM ground state and if it was indeed realized in CrO$_2$ one could hardly expect any other possibilities
for further improving the FM properties of this material. However, it is only a idealistic picture and in reality there are
always some factors, which deteriorate it. One of them is the crystal-field splitting of the $t_{2g}$ orbitals.
As somewhat extreme scenario,
one can imagine the situation when one of the $t_{2g}$ orbitals is split off by the crystal field, while all three
$t_{2g}$ the orbitals form the bands of approximately the same bandwidth (see Fig.~\ref{fig.modelDOS}b). As we will see below,
such situation could be indeed realized if, instead of CrO$_2$, one uses the experimental crystal structure of VO$_2$. In this case,
the system is close to the insulating state, and
one can expect antiferromagnetic (AFM) character of exchange interactions,
due to the superexchange mechanism~\cite{PWA}.
Nevertheless, even in this case, the FM ground state can be restored by the hole doping, which partially depopulates the
occupied bands and, thus, activates the DE mechanism. Such scenario is also frequently realized in the colossal magnetoresistive
manganites~\cite{Dagotto}. Then, even in the ideal electronic structure, depicted in Fig.~\ref{fig.modelDOS}a, one
unexplored possibility is the partial depopulation of the ``narrow'' band, formed by more localized $t_{2g}$ orbitals:
since it has finite bandwidth, it can also give rise to the DE interactions, which will compensate the decrease of the
FM interactions, cause by the depopulation of the ``wide'' bands. Thus, the situation is indeed very subtle and depends on the actual
ratio of the
bandwidths of the ``narrow'' and ``wide'' bands, as well as other interactions, including the AFM ones, existing in the system
beyond the conventional DE limit~\cite{PRL99}.

  In the present work we systematically study the doping-dependence of interatomic exchange interactions in CrO$_2$,
by using the effective low-energy electron model for the $t_{2g}$ bands, derived from the first-principles electronic
structure calculations, and solving it by means of DMFT.

  The rest of the article is organized as follows. In Sec.~\ref{sec:method}, we will briefly discuss our method of
construction of the effective low-energy electron model, the solution of this model by using the dynamical mean-field theory,
and the theory of infinitesimal spin rotations for calculations of interatomic exchange interactions~\cite{JHeisenberg,Katsnelson2000}.
Then, in Sec.~\ref{sec:results}, we will present our results. Particularly, we will discuss the doping-dependence of interatomic
exchange interactions and explain how it is related to the electronic structure of CrO$_2$ in DMFT.
After that, we will use the obtained parameters of exchange interactions
in order to evaluate the Curie temperature and
find the optimal concentration of $t_{2g}$ electrons ($n$), which maximizes $T_{\rm C}$.
We will also discuss the effect of the crystal structure on this doping dependence by considering the
experimental rutile structure of CrO$_2$ and VO$_2$ and argue that in both cases the optimal $n$ is close to $1$.
Finally, in Sec.~\ref{sec:summary}, we will present a brief summary of our work.

\section{\label{sec:method} Method}
In this section, we briefly remind the reader the main steps of our approach. More details can be found in ref.~\cite{PRB2015}.
All calculations have been performed using parameters of the experimental rutile structure
(the space group $P4_2/mnm = D_{4h}^{14}$) for CrO$_2$ \cite{Porta} and VO$_2$~\cite{McWhan}.

  The first step is the construction of the effective Hubbard-type model for the magnetically active $t_{2g}$ bands:
\begin{equation}
\hat{\cal{H}}  =  \sum_{ij} \sum_\sigma \sum_{ab}
t_{ij}^{ab}\hat{c}^\dagger_{i a \sigma}
\hat{c}^{\phantom{\dagger}}_{j b \sigma} +
  \frac{1}{2}
\sum_{i}  \sum_{\sigma \sigma'} \sum_{abcd} U^i_{abcd}
\hat{c}^\dagger_{i a \sigma} \hat{c}^\dagger_{i c \sigma'}
\hat{c}^{\phantom{\dagger}}_{i b \sigma}
\hat{c}^{\phantom{\dagger}}_{i d \sigma'},
\label{eqn.ManyBodyH}
\end{equation}
starting from the band structure in LDA. The model itself is formulated in the basis of
Wannier functions, which were obtained using the projector-operator technique~\cite{review2008,WannierRevModPhys}
and the orthonormal linear muffin-tin orbitals (LMTO's) \cite{LMTO} as the trial wave functions.
$\sigma (\sigma')$$=$ $\uparrow$/$\downarrow$ in (\ref{eqn.ManyBodyH}) are the spin indices, while $a$, $b$, $c$, and $d$ label
three $t_{2g}$ orbitals, which
have the following form in the global coordinate frame:
$|1 \rangle = \pm$$\frac{1}{2}|xy \rangle$$+$$\frac{\sqrt{3}}{2}|3z^2$$-$$r^2 \rangle$,
$|2 \rangle = \frac{1}{\sqrt{2}}|yz \rangle$$\pm$$\frac{1}{\sqrt{2}}|zx \rangle$, and
$|3 \rangle = |x^2$$-$$y^2 \rangle$,
where two signs correspond to two Cr (V) sites in the primitive cell.
Since three $t_{2g}$ orbitals belong to different irreducible representations of the point group $mmm = D_{2h}$,
all local quantities, including the DMFT self-energy and local Green's function,
are diagonal with respect to the orbital indices. The indices $i$ and $j$ specify the positions of the Cr (V) atoms in the lattice.
The parameters of the one-electron part,
$\hat{t} = [t_{ij}^{ab}]$, are defined as the matrix elements of the LDA Hamiltonian in the Wannier basis \cite{review2008}.
The parameters of screened on-site Coulomb interactions, $\hat{U} = [U^i_{abcd}]$, are calculated in the framework of
constrained random-phase approximation (RPA) \cite{Ferdi04}, using the simplified procedure, which is explained in ref.~\cite{review2008}.
The parameters of the model Hamiltonian, obtained for CrO$_2$ and VO$_2$, are summarized in Table~\ref{tab.parameters}.
\begin{table}[h!]
\caption{\label{tab.parameters}
Parameters of electron model (in eV) for CrO$_2$ and VO$_2$:
the crystal-field splitting of the $t_{2g}$ levels ($\hat{t}_{00}$),
the transfer integrals between nearest- and next-nearest neighbors ($\hat{t}_{01}$ and $\hat{t}_{02}$, respectively),
and Kanamori parameters of the intra-orbital Coulomb (${\cal U}$) and exchange (${\cal J}$) interactions.
The positions of the atomic sites $1$ and $2$ relative to the origin are explained in Fig.~\protect\ref{fig.CrO2Ji}.}
\begin{indented}
\item[]\begin{tabular}{@{}ccc}
\br
  & CrO$_2$ & VO$_2$ \\
\mr
$\hat{t}_{00}$
&
$\left(
\begin{array}{ccc}
 -0.246 &    0    &   0   \\
    0   &  0.060  &   0   \\
    0   &    0    & 0.186
\end{array}
\right)$
&
$\left(
\begin{array}{ccc}
 -0.225 &    0   &   0   \\
    0   & -0.081 &   0   \\
    0   &    0   & 0.306
\end{array}
\right)$
\\
$\hat{t}_{01}$
&
$\left(
\begin{array}{ccc}
 -0.067 &    0   &   0   \\
    0   & -0.191 &   0   \\
    0   &    0   & 0.158
\end{array}
\right)$
&
$\left(
\begin{array}{ccc}
 -0.212 &    0   &   0   \\
    0   & -0.123 &   0   \\
    0   &    0   & 0.200
\end{array}
\right)$
\\
$\hat{t}_{02}$
&
$\left(
\begin{array}{ccc}
 -0.015 &   0   &    0   \\
 -0.028 &   0   &    0   \\
    0   & 0.194 & -0.119
\end{array}
\right)$
&
$\left(
\begin{array}{ccc}
 \phantom{-}0.004 &   0   &   0    \\
 -0.059 &   0   &   0    \\
    0   & 0.178 & -0.127
\end{array}
\right)$
\\
${\cal U}$ &  $2.84$   &  $0.70$  \\
${\cal J}$ &  $2.98$   &  $0.71$  \\
\br
\end{tabular}
\end{indented}
\end{table}
\begin{figure}[h!]
\begin{center}
\resizebox{15cm}{!}{\includegraphics{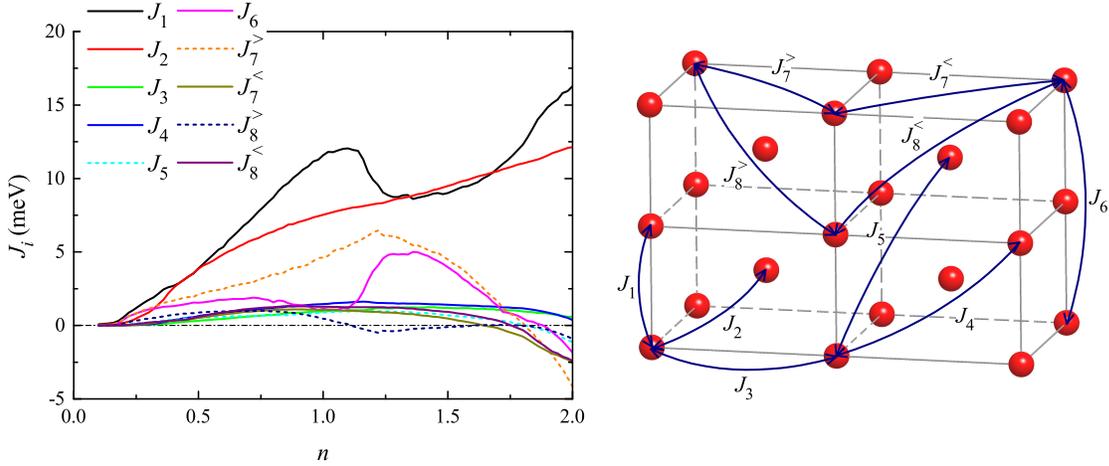}}
\end{center}
\caption{\label{fig.CrO2Ji} (Left) Dependence of interatomic exchange interactions on the
number of electrons in the $t_{2g}$ band of CrO$_2$.
(Right) Lattice of Cr sites with the notation of interatomic exchange interactions.}
\end{figure}

  After the construction, the model (\ref{eqn.ManyBodyH}) was solved by means of the DMFT \cite{DMFTRevModPhys},
where the properties of the many-electron system are formally related to the
frequency-dependent
self-energy $\hat{\Sigma}^{\uparrow, \downarrow}(\omega)$, which describes all kind of correlation effects in
the one-electron Green's function $\hat{G}^{\uparrow, \downarrow}(\omega, {\bf k})$ (in the reciprocal ${\bf k}$-space):
\begin{eqnarray}
\hat{G}^{\uparrow, \downarrow}(\omega, {\bf k}) =
\left[\omega - \hat{t}({\bf {k}}) - \hat{\Sigma}^{\uparrow, \downarrow}(\omega) \right]^{-1}.
\label{eqn.latticeG}
\end{eqnarray}
The basic approximation here is that $\hat{\Sigma}^{\uparrow, \downarrow}(\omega)$ does not depend on ${\bf k}$, which becomes
exact in the limit of infinite dimensions (or coordination numbers).
The main idea of DMFT is to map the initial many-body problem for the crystalline lattice onto the quantum impurity one,
surrounded by an effective electronic bath, and find parameters of Green's function of this bath,
${\cal G}^{\uparrow, \downarrow}(\omega)$, by solving the Anderson impurity model. For these purposes we employ
the newly developed exact diagonalization scheme. The details can be found in refs.~\cite{PRB2015,arxiv}.
Then, $\hat{\Sigma}^{\uparrow, \downarrow}(\omega)$ is obtained from the Dyson equation:
\begin{eqnarray}
\hat{G}^{\uparrow, \downarrow}(\omega) =
\hat{\cal G}^{\uparrow, \downarrow}(\omega) +
\hat{\cal G}^{\uparrow, \downarrow}(\omega) \hat{\Sigma}^{\uparrow, \downarrow}(\omega) \hat{G}^{\uparrow, \downarrow}(\omega),
\label{eqn.Dyson}
\end{eqnarray}
where $\hat{G}^{\uparrow, \downarrow}(\omega) = \sum_{{\bf k}} \hat{G}^{\uparrow, \downarrow}(\omega, {\bf k})$,
and the problem is solved self-consistently.

  After the solution of the DMFT problem, we consider the mapping of the electron model (\ref{eqn.ManyBodyH}) onto the
spin Heisenberg model:
\begin{equation}
\hat{\cal{H}}_S = -\frac{1}{2} \sum_{ij} J_i \hat{\boldsymbol{e}}_{j} \cdot \hat{\boldsymbol{e}}_{j+i}.
\label{eqn.SH}
\end{equation}
In these notations, $J_i$ is the exchange coupling between two Cr (V) sites, located in the origin ($0$)
and in the point $i$ of the lattice, relative to the origin, and $\hat{\boldsymbol{e}}_{j}$ is the direction of the
spin moment at the site $j$ (while the value of the spin itself is included in the definition of $J_i$).
The mapping onto the spin model implies the adiabatic motion of spins when all instantaneous changes of the
electronic structure adjust slow rotations of the spin magnetic moments.
Then, the parameters of this model can be obtained by
using the theory of infinitesimal spin rotations near the FM state~\cite{JHeisenberg,Katsnelson2000}:
\begin{equation}
J_i = \frac{1}{2\pi} {\rm Im} \int_{- \infty}^{\varepsilon_{\rm F}} d \omega \, {\rm Tr}_L \left\{
\Delta \hat{\Sigma}(\omega) \hat{G}_{0i}^{\uparrow}(\omega)
\Delta \hat{\Sigma}(\omega) \hat{G}_{i0}^{\downarrow}(\omega)
\right\},
\label{eqn:Jij}
\end{equation}
where $\hat{G}^{\uparrow, \downarrow}_{0i}(\omega) = [ \omega - \hat{t} -
\hat{\Sigma}^{\uparrow, \downarrow}(\omega) ]^{-1}_{0i}$
is the one-electron Green function between sites $0$ and $i$,
$\Delta \hat{\Sigma} = \hat{\Sigma}^\uparrow - \hat{\Sigma}^\downarrow$, and ${\rm Tr}_L$ denotes the trace over the orbital indices.
In fact, the parameters $\{ J_i \}$ given by Eq.~(\ref{eqn:Jij}) are proportional to the second derivatives
of the total energy with respect to the infinitesimal rotations of spins~\cite{JHeisenberg,Katsnelson2000}.
Thus, these parameters characterize the local stability of the
magnetic state.

\section{\label{sec:results} Results and Discussions}
\begin{figure}[h!]
\begin{center}
\resizebox{15cm}{!}{\includegraphics{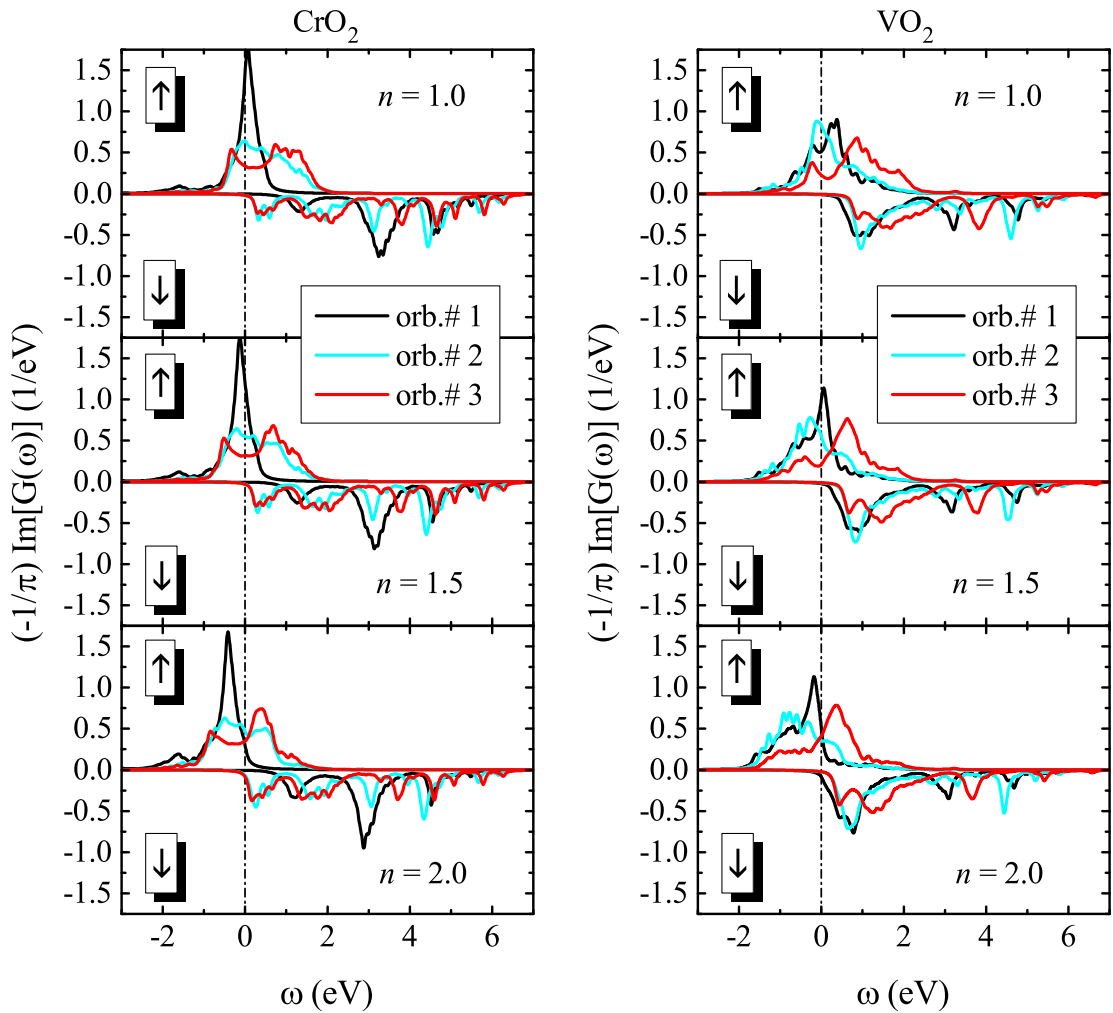}}
\end{center}
\caption{\label{fig.DOS} Partial densities of states as obtained in the DMFT calculations
for CrO$_2$ (left) and VO$_2$ (right) for $n=$ $1.0$, $1.5$, and $2.0$ electrons in the $t_{2g}$ band.
The Fermi level is at zero energy (shown by dot-dashed line).}
\end{figure}

\begin{figure}[h!]
\begin{center}
\resizebox{8cm}{!}{\includegraphics{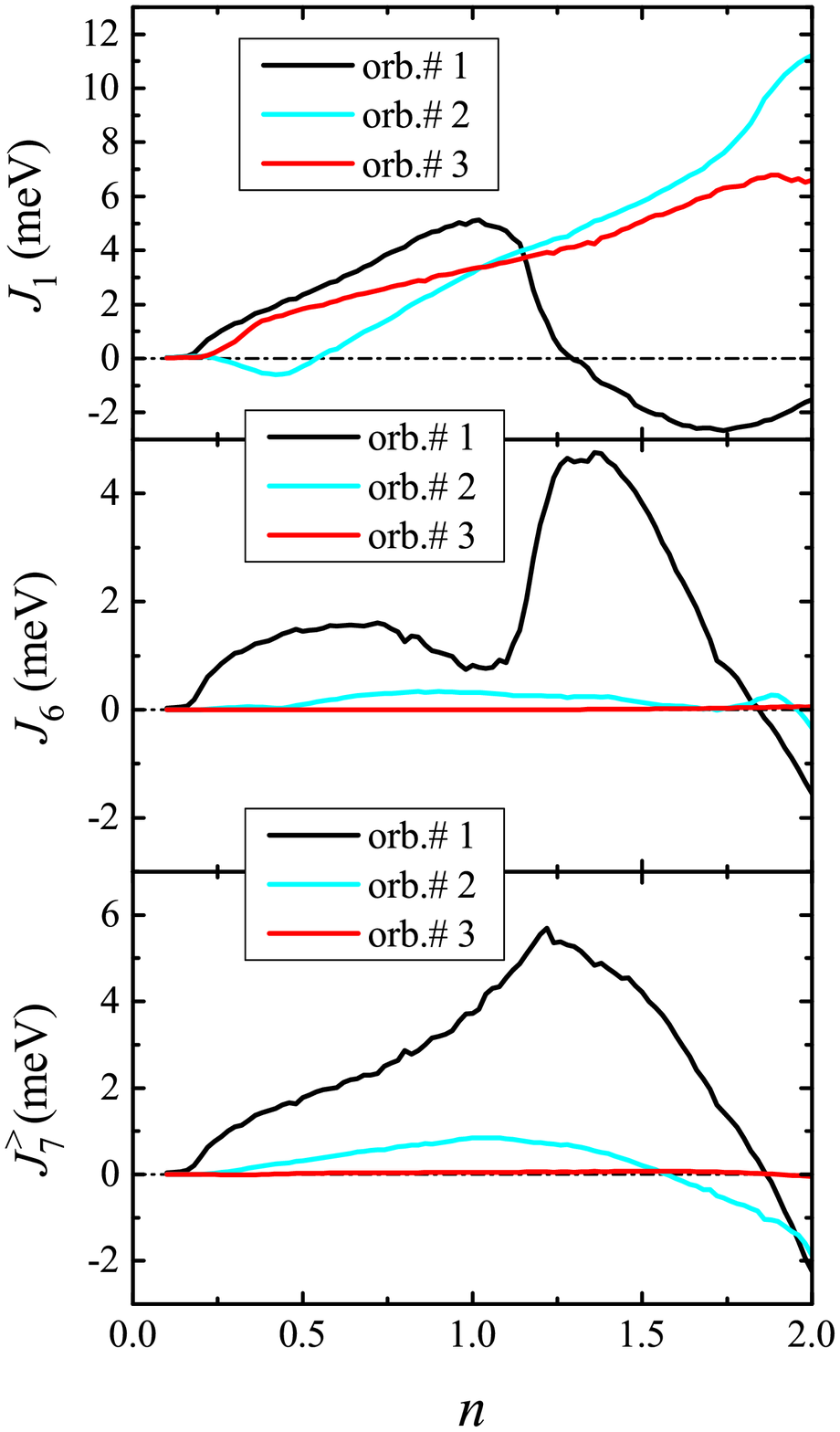}}
\end{center}
\caption{\label{fig.Jpartial} Partial contributions of three $t_{2g}$ orbitals to the
nearest-neighbor exchange coupling $J_1$ and long-range interactions $J_6$ and $J_7^>$ in CrO$_2$.}
\end{figure}

First, let us explain the behavior of nearest-neighbor (NN) exchange interaction, $J_1$, which for CrO$_2$ exhibits a
nonmonotonous behavior with a peak at around $n=1$ (see Fig.~\ref{fig.CrO2Ji}).
Due to the $mmm$ symmetry of the NN bond $\langle 01 \rangle$, the matrices of transfer integrals, $\hat{t}_{01}$, and
Green's function, $\hat{G}_{01}$, are diagonal with respect to the orbital indices. Therefore, $J_1$ is the superposition
of three individual contributions originating from the three $t_{2g}$ orbitals. Due to the DE mechanism,
which largely controls the behavior of $J_1$, each contribution is expected to have a maximum near the half-filling and then decrease
with the decrease of $n$~\cite{PRL99}.
Therefore, near $n=2$, where the $\uparrow$-spin band,
formed by the orbital $1$, is nearly fully populated while the bands $2$ and $3$ are approximately half-filled
(see Fig.~\ref{fig.DOS}), $J_1$ decrease with the decrease of $n$.
However, then the orbital $1$ is gradually depopulated and
becomes half-filled at around $n=1$. This explain the appearance of the additional peak around $n=1$ (Fig.~\ref{fig.Jpartial}).
Indeed, the transfer integral between orbitals of the type $1$
in the bond $\langle 01 \rangle$ is moderately strong
(see Table~\ref{tab.parameters})~\cite{PRB2015}.
Therefore, one can expect relatively large FM DE contribution and weak (in comparison with other orbitals)
AFM contribution to the exchange coupling,
which appears in the next order of the $(\Delta \hat{\Sigma})^{-1}$ expansion~\cite{PRB2015,PRL99}.
The $t_{2g}$ electrons, occupying the orbital $1$ in CrO$_2$, are sometimes considered as the localized ones
and participating only in the formation of the
weakly AFM background (similar to $t_{2g}$ electrons in manganites~\cite{Dagotto}), while the ferromagnetism is entirely related to the
itinerant electrons in the bands,
formed by the orbitals $2$ and $3$~\cite{Korotin}. As we have seen , such interpretation is not quiet correct
and the orbital $1$ can also substantially contribute to the ferromagnetism, at least in the case of the hole doping.

  For the next-NN bond $\langle 02 \rangle$, the transfer integrals are strongly off-diagonal with respect to the
orbitals indices
(see Table~\ref{tab.parameters}).
Moreover, the matrix element $t_{02}^{11}$ is small. Therefore, the contribution of the orbital $1$ to $J_2$ will be also
small and entangled with other orbitals. In such situation, the behavior of $J_2$ is mainly controlled by the
population of the orbitals $2$ and $3$.
This is the main reason why $J_2$ monotonously decrease with the decrease of $n$.

  The behavior of longer-range interactions is more complex. Very generally, one can expect that the more remote are the sites,
the more oscillations will have the exchange coupling between them depending on $n$~\cite{Heine1,Heine2}.
This tendency is indeed observed in our DMFT calculations (see Fig.~\ref{fig.CrO2Ji}): while $J_1$ and $J_2$ remain FM for $n \le 2$,
several long-range interactions change their sign. Particularly, the interactions $J_5$-$J_8$ are AFM at $n=2$, but
gradually become FM with the decrease of $n$. The strongest effect is observed for $J_6$ and $J_7^>$, which reach the maximum
at around $n = 1.25$. This behavior is also related to the partial depopulation of the $\uparrow$-spin band, formed by the orbital $1$,
while the contribution of two other orbitals is considerably weaker (see Fig.~\ref{fig.Jpartial}).

\begin{figure}[h!]
\begin{center}
\resizebox{15cm}{!}{\includegraphics{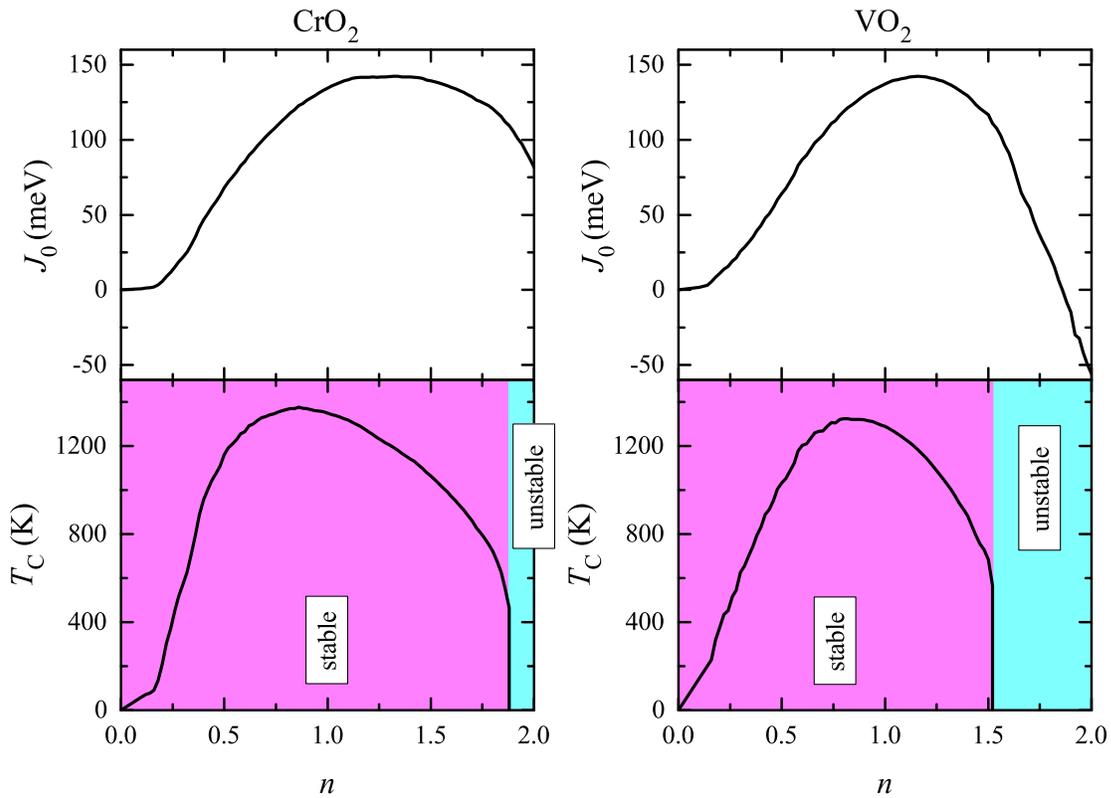}}
\end{center}
\caption{\label{fig.Tc}
Total exchange coupling around each transition-metal site ($J_0$)
and the Curie temperature ($T_{\rm C}$), evaluate in the framework of Tyablikov's random-phase approximation,
depending on the number of $t_{2g}$ electrons in CrO$_2$ (left) and VO$_2$ (right).}
\end{figure}

  Thus, while for $n=2$ the FM state in our model is unstable (see Fig.~\ref{fig.Tc}),
one can expect a strong ferromagnetism around $n=1$, mainly as the joint
effect of $J_1$, $J_6$, and $J_7^>$. Note that in our model, we do not consider the contributions caused by direct exchange interactions and
the polarization of the oxygen band, which are indeed crucial for stabilizing the FM ground state in CrO$_2$~\cite{PRB2015}. However, it is
reasonable to expect that the doping-dependence of these interactions is relatively weak,
because the doping mainly affects the magnetic
interactions in the $t_{2g}$ band. Therefore, we do not consider here the direct exchange interactions as well as
the effects of magnetic polarization of the oxygen band, assuming that, if necessary, these
(independent on the doping) FM contributions can be simply added to the ones in the $t_{2g}$ band, which we explore in the
present work.

  In order to discuss the
band-filling dependence of $T_{\rm C}$, we evaluate two quantities, which are related to it. The first one is the sum
of exchange interactions around some central site, $J_0 = \sum_i J_i$, which is proportional to the mean-field estimate for $T_{\rm C}$.
The second one is the actual value of $T_{\rm C}$ for the spin model (\ref{eqn.SH}),
evaluated using Tyablikov's RPA \cite{spinRPA}.
The results are explained in Fig.~\ref{fig.Tc}.
The definition of $T_{\rm C}$ is only meaningful when the FM state is stable. In Tyablikov's approach, this is defined by the
magnon frequencies $\omega({\bf q})$: if some of $\omega({\bf q})$ are negative,
the FM state is considered to be unstable. These regions of
stability (and instability) of the FM state are also shown in Fig.~\ref{fig.Tc}. $J_0$ and $T_{\rm C}$ reach the maximums at
around $n=$ $1.3$ and $0.9$, respectively. The maximal value of $T_{\rm C}$ is about $1400$ K, which is indeed very high.

\begin{figure}[h!]
\begin{center}
\resizebox{10cm}{!}{\includegraphics{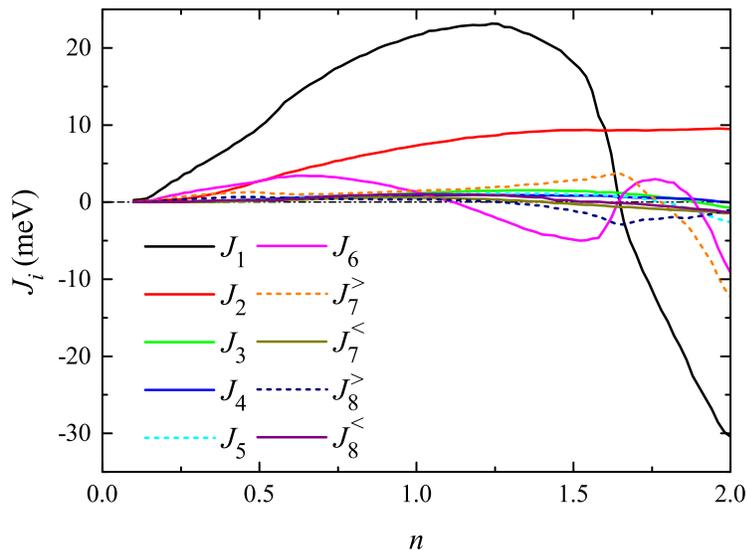}}
\end{center}
\caption{\label{fig.VO2Ji} Dependence of interatomic exchange interactions on the
number of electrons in the $t_{2g}$ band of VO$_2$.}
\end{figure}

  Finally, we discuss the effect of the crystal structure on the considered properties. For these purposes, instead of CrO$_2$,
we take the experimental crystal structure of VO$_2$ and perform the same calculations.
Indeed, if the optimal band-filling in CrO$_2$ is around $n=1$, one can also try to consider the possibility of the
electron or hole doping of VO$_2$.
The obtained band-filling dependence of interatomic exchange interactions in VO$_2$ is explained in Fig.~\ref{fig.VO2Ji} and
corresponding densities of states are shown in Fig.~\ref{fig.DOS}.
Judging from the parameters of the model Hamiltonian (see Table~\ref{tab.parameters}), the main difference between CrO$_2$ and VO$_2$
is in the transfer integrals operating in the NN bond $\langle 01 \rangle$: in CrO$_2$,
the transfer integrals between orbitals of the type $1$
are considerably weaker than those between orbitals $2$ and $3$, while in VO$_2$
all three types are comparable. This explains the fact that all three $\uparrow$-spin bands, formed by the orbitals $1$-$3$, are comparable
in the case of VO$_2$ (see Fig.~\ref{fig.DOS}), while in CrO$_2$ the band $1$ is considerably narrower. Moreover, the crystal field,
separating the orbital $3$, is considerably stronger in the case of VO$_2$. Therefore, for $n=2$ in VO$_2$,
the $\uparrow$-spin bands $1$ and $2$ are nearly filled (or half-filled in the non-spin-polarized case),
while the band $3$ is almost empty. This explain the AFM character of NN interaction $J_1$ at $n=2$,
as the general tendency towards antiferromagnetism near the half-filling
(referring to the bands $1$ and $2$ in the non-spin-polarized case) \cite{Heine1,Heine1}. However,
the decrease of $n$ leads to the partial depopulation of the $\uparrow$-spin bands $1$ and $2$, which changes the
character of the NN interaction $J_1$ towards the FM one
due to the DE mechanism. This FM coupling is especially strong around $n=1.25$. The relative strength of long-range
interactions is considerably weaker. Thus these interactions play a less important role in comparison with CrO$_2$.

  The mean-field Hartree-Fock (HF) calculations, performed for the FM state of VO$_2$,
yield $J_1 = 25.2$ meV and $J_2 = 7.5$ meV. These values are somewhat larger than the ones obtained in the DMFT
($21.6$ and $7.3$ meV, respectively) due to the lack of dynamic correlations, which effectively decrease
the intraatomic exchange splitting between majority- and minority-spin states near the Fermi level and, thus,
additionally strengthens the AFM superexchange interactions in the case of DMFT~\cite{PRB2015}. Nevertheless,
both HF calculations and DMFT support strong FM character of exchange interactions in
VO$_2$, if they are calculated near the FM state.
It is also important to note that the electronic structure and
exchange interactions in VO$_2$ exhibit the strong dependence
on the magnetic state in which they are calculated (and in this sense, VO$_2$ can be regarded as a
\textit{smart material}~\cite{Nori,Molaei}).
For instance, in the HF calculations,
the AFM alignment of two V spins in the
primitive cell opens the band gap (of about $0.1$ eV) and changes the character of NN interactions to
the AFM one ($J_1 = -$$14.6$ meV). This AFM configuration is higher in energy (and, thus, cannot be realized as the ground state).
However, it can contribute to the thermodynamic average at elevated temperatures. Thus, the FM ground state obtained
for the rutile structure of VO$_2$ is not inconsistent with the negative Curie-Weiss temperature,
expected from the high-temperature behavior of magnetic susceptibility~\cite{VO2magneticexp}: simply, the
electronic structure of the paramagnetic state should be very different from the one realized in the FM ground state
and, therefore, the magnetic interactions should be also different.

  Using parameters of exchange interactions, obtained in DMFT,
we again evaluate the ``mean-field parameter'' $J_0$ and
the Curie temperature, using Tyablikov's RPA. The results are summarized in Fig.~\ref{fig.Tc}.
Despite substantial differences in the electronic structure (see Fig.~\ref{fig.DOS}), the Curie temperature
exhibits very similar behavior as the function of the band-filling in CrO$_2$ and VO$_2$. In the latter case,
$J_0$ and $T_{\rm C}$ take the maximum at around $n=$ $1.2$ and $0.8$, respectively. The maximal value of $T_{\rm C}$
is close to $1400$ K, i.e. similar to CrO$_2$.

\section{\label{sec:summary} Summary and Conclusions}
By combining first-principles electronic structure calculations with the dynamical mean-field theory, we have
investigated the band-filling dependence of interatomic exchange interactions and the Curie temperature in CrO$_2$ and
related rutile compounds.
We have argued that $T_{\rm C}$ in the rutile compounds can be substantially increased
by controlling the number of the $t_{2g}$ electrons.
It appears that, as far as the Curie temperature is concerned,
the filling of the $t_{2g}$ band,
which is realized in the stoichiometric CrO$_2$, is not the most optimal one
and much higher $T_{\rm C}$ can be expected for the smaller number of
the $t_{2g}$ electrons due to the general tendency
of narrow-band compounds towards the ferromagnetism at the beginning and the end of the band-filling~\cite{Heine1,Heine2}.
According to our estimates, the optimal number of $t_{2g}$ electrons should be close to $n=1$,
which formally corresponds to VO$_2$. Thus, if VO$_2$ indeed crystallized in the rutile structure, it could be a good
FM material with some of the properties even exceeding CrO$_2$. However, below $340$ K, VO$_2$ undergoes the
transition to the monoclinic phase, which is accompanied by the dimerization of the V sites and the formation of the nonmagnetic
insulating state~\cite{VO2magneticexp,Morin,Belozerov}.
Actually, the situation when the robust ferromagnetism, expected in some materials,
is destroyed by the lattice distortion is rather common in nature. One of the well known examples is LaMnO$_3$,
which from the viewpoint of the DE physics is expected to be a strong ferromagnet with the optimal
concentration of electrons in the partially filled $e_g$ band~\cite{PRL99}.
However, this FM state in LaMnO$_3$ is destroyed by the Jahn-Teller distortion, which is responsible
for the formation of the layered AFM state~\cite{PRL96}. Nevertheless, the FM state can be realized in the thin films,
where the crystal structure is controlled by the substrate.
Such a possibility for VO$_2$ was indeed demonstrated in refs.~\cite{Nori,Molaei}, where it was argued that by
deposing VO$_2$ on different substrates one can achieve the
room-temperature ferromagnetism, which can be switched `on' and `off', depending on the treatment conditions.
This behavior was attributed to the V$^{3+}$ defects, existing in the thin films of VO$_2$.
However, results of our work demonstrate that such behavior can be also intrinsic and related to the
general tendency towards the ferromagnetism in the narrow-band compounds at the beginning of the band filling.

\ack
This work is partly supported by the grant of Russian Science Foundation (project No. 14-12-00306).


\section*{References}
\begin{thebibliography}{99}

\bibitem{SkomskiCoey}
Skomski R and Coey J M D  1999
\textit{Permanent Magnetism}
(New York: Taylor \& Francis Group)

\bibitem{Skomski}
Skomski R 2008
\textit{Simple Models of Magnetism}
(Oxford: Oxford University Press)

\bibitem{Schwarz}
Schwarz K 1986
\textit{J. Phys. F: Met. Phys.} \textbf{16} L211

\bibitem{Mazin}
Mazin I I, Singh D J and Ambrosch-Draxl C 1999
\textit{Phys. Rev. B} \textbf{59} 411

\bibitem{Chioncel}
Chioncel L, Allmaier H, Arrigoni E, Yamasaki A, Daghofer M, Katsnelson M I and Lichtenstein A I 2007
\textit{Phys. Rev. B} \textbf{75} 140406(R)

\bibitem{HMRevModPhys}
Katsnelson M I, Irkhin V Yu, Chioncel L, Lichtenstein A I and de Groot R A 2008
\textit{Rev. Mod. Phys.} \textbf{80} 315

\bibitem{Singh}
Singh A, Voltan S, Lahabi K and Aarts J 2015
\textit{Phys. Rev. X} \textbf{5} 021019

\bibitem{PRB2015}
Solovyev I V, Kashin I V and Mazurenko V V 2015
\textit{Phys. Rev. B} \textbf{92} 144407

\bibitem{DMFTRevModPhys}
Georges A, Kotliar G, Krauth W and Rozenberg M J 1996
\textit{Rev. Mod. Phys.} \textbf{68} 13

\bibitem{PRL99}
Solovyev I V and Terakura K 1999
\textit{Phys. Rev. Lett.} \textbf{82} 2959

\bibitem{Dagotto}
Dagotto E, Hotta T and Moreo A  2001
\textit{Phys. Rep.} \textbf{344} 1

\bibitem{Korotin}
Korotin M A, Anisimov V I, Khomskii D I and Sawatzky G A 1998
\textit{Phys. Rev. Lett.} \textbf{80} 4305

\bibitem{PWA}
Anderson P W 1959
\textit{Phys. Rev.} \textbf{115} 2

\bibitem{JHeisenberg}
Liechtenstein A I, Katsnelson M I, Antropov V P and Gubanov V A 1987
\textit{J. Magn. Magn. Mater.} \textbf{67} 65

\bibitem{Katsnelson2000}
Katsnelson M I and Lichtenstein A I 2000
\textit{Phys. Rev. B} \textbf{61} 8906

\bibitem{Porta}
Porta P, Marezio M, Remeika J P and Dernier P D 1972
\textit{Mater. Res. Bull.} \textbf{7} 157

\bibitem{McWhan}
McWhan D B, Marezio M, Remeika J P and Dernier P D 1974
\textit{Phys. Rev. B} \textbf{10} 490

\bibitem{review2008}
Solovyev I V 2008
\textit{J. Phys.: Condens. Matter} \textbf{20} 293201

\bibitem{WannierRevModPhys}
Marzari N, Mostofi A A, Yates J R, Souza I and Vanderbilt D 2012
\textit{Rev. Mod. Phys.} \textbf{84} 1419

\bibitem{LMTO}
Gunnarsson O, Jepsen O and Andersen O K 1983
\textit{Phys. Rev. B} \textbf{27} 7144

\bibitem{Ferdi04}
Aryasetiawan F, Imada M, Georges A, Kotliar G, Biermann S and Lichtenstein A I 2004
\textit{Phys. Rev. B} \textbf{70} 195104

\bibitem{arxiv}
Kashin I V and Mazurenko V V 2015
\textit{arXiv}:1508.04895

\bibitem{Heine1}
Heine V and Samson J H 1980
\textit{J. Phys. F: Metal Phys.} \textbf{10} 2609

\bibitem{Heine2}
Heine V and Samson J H 1983
\textit{J. Phys. F: Metal Phys.} \textbf{13} 2155

\bibitem{spinRPA}
Tyablikov S V 1975 \textit{Methods of Quantum Theory of Magnetism}
(Moscow: Nauka)

\bibitem{Nori}
Nori S, Yang T-H and Narayan J 2011
\textit{Magnetic Materials and Devices} \textbf{63} 29

\bibitem{Molaei}
Molaei R, Bayati R, Nori S, Kumar D, Prater J T and Narayan J 2013
\textit{Appl. Phys. Lett.} \textbf{103} 252109

\bibitem{VO2magneticexp}
Pouget J P, Lederer P, Schreiber D S, Launois H, Wohlleben D, Casalot A and Villeneuve G 1972
\textit{J. Phys. Chem. Solids.} \textbf{33} 1961

\bibitem{Morin}
Morin F J 1959
\textit{Phys. Rev. Lett.} \textbf{3} 34

\bibitem{Belozerov}
Belozerov A S, Korotin M A, Anisimov V I and Poteryaev A I 2012
\textit{Phys. Rev. B} \textbf{85} 045109

\bibitem{PRL96}
Solovyev I, Hamada N and Terakura K 1996
\textit{Phys. Rev. Lett.} \textbf{76} 4825


\end{thebibliography}
\end{document}